\documentclass[9pt,letterpaper,english,conference]{IEEEtran}
\usepackage[T1]{fontenc}
\usepackage[latin9]{inputenc}
\usepackage{color}
\usepackage{units}
\usepackage{amsmath}
\usepackage{graphicx}
\usepackage[numbers]{natbib}

\makeatletter

\providecommand{\tabularnewline}{\\}

\makeatletter
\IEEEoverridecommandlockouts
\def\ps@headings{%
\def\@oddhead{\mbox{}\scriptsize\rightmark \hfil \thepage}%
\def\@evenhead{\scriptsize\thepage \hfil \leftmark\mbox{}}%
\def\@oddfoot{}%
\def\@evenfoot{}}
\makeatother
\makeatother

\usepackage{babel}

\begin{document}

\title{\textcolor{black}{Co-Designing Multi-Packet Reception, Network Coding,
and MAC Using a Simple Predictive Model\thanks{This work is sponsored by the Office of the Secretary of Defense  Contract FA8721-05-C-0002. Opinions, interpretations, recommendations, and conclusions are those of the authors and are not necessarily endorsed by the United States Government. Specifically, this work was supported by Information Systems of D,DR\&E. Contributions of the Irwin Mark Jacobs and Joan Klein Jacobs Presidential Fellowship have also been critical to the success of this project.}}}

\author{\textcolor{black}{Jason Cloud{*}$^{\dagger}$, Linda Zeger$^{\dagger}$,
Muriel Médard{*}}\\
\textcolor{black}{{*}Research Laboratory of Electronics, Massachusetts
Institute of Technology, Cambridge, MA.}\\
\textcolor{black}{$^{\dagger}$MIT Lincoln Laboratory, Lexington,
MA.}\\
\textcolor{black}{Email: \{jcloud@, zeger@ll., medard@\}mit.edu}}
\maketitle
\begin{abstract}
\textcolor{black}{We design a cross-layer approach to optimize the
joint use of multi-packet reception and network coding, in order to
relieve congestion. We construct a model for the behavior of the 802.11
MAC and apply it to several key canonical topology components and
their extensions to any number of nodes. The results obtained from
this model match the available experimental results, which are for
routing and opportunistic network coding, with fidelity. Using this
model, we show that fairness allocation by the MAC can seriously impact
performance; hence, we devise a new MAC that not only substantially
improves throughput relative to the current 802.11 MAC, but also provides
fairness to }\textit{\textcolor{black}{flows}}\textcolor{black}{{} of
information rather than to }\textit{\textcolor{black}{nodes}}\textcolor{black}{.
We show that the proper combination of network coding, multi-packet
reception, and our new MAC protocol achieves super-additive throughput
gains of up to 6.3 times that of routing alone with the use of the
standard 802.11 MAC. Finally, we extend the model to analyze the asymptotic
behavior of our new MAC as the number of nodes increases.}
\end{abstract}
\textcolor{black}{\IEEEpeerreviewmaketitle}

\section{\textcolor{black}{Introduction}}

\textcolor{black}{With the increase in wireless use, current wireless
systems are throughput limited and are difficult to scale to large,
dense networks. We develop a simple model that is easily extended
to analyze the asymptotic regime so that we can evaluate the performance
of combining various techniques to increase network throughput and
reduce overall delay.}

\begin{figure}
\begin{centering}
\textcolor{black}{\includegraphics[width=3.25in]{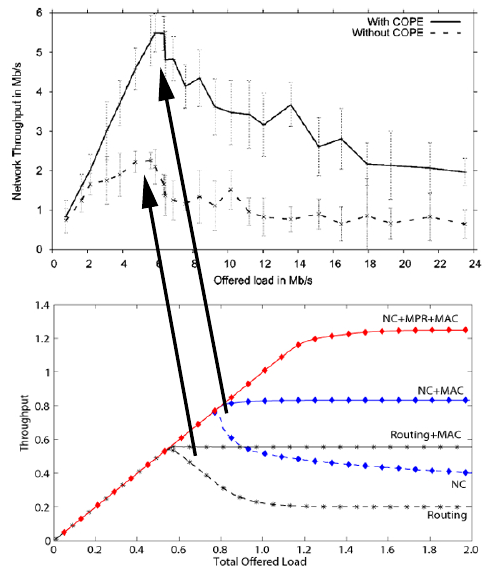}}
\par\end{centering}

\textcolor{black}{\caption{Comparison of the empirical COPE performance data collected from a
20-node 802.11 wireless ad hoc network test bed (top), \citep{Katti00},
and the resulting throughput using a model of the 802.11 MAC proposed
by \citep{Zhao00} (bottom). This model is th\textcolor{black}{e starting
point for o}ur analysis with MPR and development of our improved MAC.
\label{fig:Emperical-COPE-performance}}
}
\end{figure}

\textcolor{black}{The introduction of network coding \citep{Ahlswede00}
led to the proposal of a new forwarding architecture, COPE, for wireless
networks. Proposed by Katti }\textsl{\textcolor{black}{et al.}}\textcolor{black}{{}
\citep{Katti00}, COPE identifies coding opportunities and exploits
them by forwarding multiple packets in a single transmission. The
use of this simple coding scheme was shown to provide up to 3 to 4
times the throughput capacity. Implementing COPE in a 20-node 802.11
test bed, Katti }\textit{\textcolor{black}{et al}}\textcolor{black}{.
provided empirical data, seen in the upper half of Fig. \ref{fig:Emperical-COPE-performance},
which shows the benefits of using COPE in wireless mesh networks.}%

Sengupta \textit{et al.}, \citep{4215706} and Le \textit{et al.},
\citep{4509678} provided analyses of these experimental results,
but only considered coding a maximum of two packets together at a
time or did not address the interaction between network coding and
the MAC fairness. \textcolor{black}{As a result, their analyses provide
throughput gains that are considerably smaller than the experimental
results and do not explain the non-monotonic behavior of the experimental
results in Fig. \ref{fig:Emperical-COPE-performance}. Zhao and Médard,
\citep{Zhao00}, modeled the same experimental results showing that
the }\textit{\textcolor{black}{fairness}}\textcolor{black}{{} imposed
by the 802.11 MAC explains this non-monotonic behavior. In addition,
they demonstrated that the majority of the throughput gain achieved
by using COPE is a result of coding three or more uncoded, or native,
packets together at time. They showed that these gains are not reflected
in three node network models and at least five nodes are required
to accurately capture the throughput effects resulting from network
coding. The network coding (NC) and routing curves in Fig. \ref{fig:Emperical-COPE-performance}
show that the results obtained using their model for a simple 5-node
cross topology component \citep{Zhao00} is consistent with the empirical
data from \citep{Katti00}. Furthermore, Seferoglu }\textit{\textcolor{black}{et.
al.}}\textcolor{black}{{} \citep{5487689} used this 5-node topology
component, and variants of them, to analyze TCP performance over coded
wireless networks. With this in mind, we consider the 5-node cross
topology component and additional 5-node topology components, as well
as their extensions to any number of nodes, in order to help in our
understanding of the effects of combining network coding and MPR in
larger networks.}

\textcolor{black}{While the performance of COPE significantly increases
network throughput \citep{Katti00}, it does not completely alleviate
the limitation of multi-user interference. With the development of
new radio technologies such as OFDMA, the ability to receive multiple
packets simultaneously makes it possible to increase throughput and
also has the potential to reduce contention among users \citep{Garcia00}.
Extensive research has been conducted on MPR with uncoded traffic.
For instance, the stability of slotted ALOHA with MPR was studied
by \citep{1272} and several protocols implementing MPR have been
proposed by \citep{1180550} and \citep{5196676}. However, little
analysis has been performed in evaluating schemes involving both MPR
and network coding. Garcia-Luna-Aceves}\textit{\textcolor{black}{{}
et al.}}\textcolor{black}{{} \citep{1280960} }\textit{\textcolor{black}{compared}}\textcolor{black}{{}
the use of network coding to MPR, but did not consider the }\textit{\textcolor{black}{combined}}\textcolor{black}{{}
use of both MPR and network coding. In addition, Rezaee }\textit{\textcolor{black}{et
al.}}\textcolor{black}{{} \citep{Rezaee(00)}, provided an analysis
of the combined use of network coding and MPR in a fully connected
network, but did not consider the effects of bottlenecks or multi-hop
traffic.}

\textcolor{black}{We instead provide an analysis of the combined use
of network coding and MPR in a multi-hop, congested network. We extend
the initial model proposed by \citep{Zhao00} to include various topology
configurations, asymptotic behavior, and MPR in order to show that
the achievable throughput when using network coding in conjunction
with MPR in }\textit{\textcolor{black}{multi-hop}}\textcolor{black}{{}
networks is super-additive. We then use this model to design a cross-layer
solution to optimize the throughput subject to constraints requiring
fairness between flows, rather than between nodes, for network structures
that induce congestion. While MAC fairness has been studied \citep{4622754},
our solution takes into account the interaction among MPR, network
coding, and MAC. Using our simplified model, we then analyze the behavior
of our solution in the asymptotic regime as the number of nodes in
each topology component increases.}

\textcolor{black}{The remainder of the paper is organized as follows:
Section \ref{sec:Network-models} describes the network models used
in our analysis. Section \ref{sec:Multi-Packet-Reception-and} provides
an analysis of network coding and MPR for 5-node network topology
components using the existing 802.11 MAC. Section \ref{sec:Improving-the-MAC}
demonstrates the importance of considering the MAC when using a combined
MPR and network coding solution and provides an improved MAC that
optimizes throughput subject to flow constraints, MPR and network
coding. Section \ref{sec:Performance-of-large-N} shows that MPR and
network coding provide significant gains when considering delay in
the asymptotic regime. Finally in Section \ref{sec:Conclusion}, we
conclude with a comparison of the results.}

\section{\textcolor{black}{Network models and parameters\label{sec:Network-models}}}

\textcolor{black}{We use a simple implementation of opportunistic
network coding, COPE \citep{Katti00}. COPE uses the broadcast nature
of the wireless channel to overhear transmissions from a node's neighbor
to extract information from any coded packet that it receives. An
example of the procedure used by COPE is seen from a 3-node tandem
network with node $R$ connected to both node $A$ and node $B$,
but $A$ is not connected to $B$. Source $A$ sends packet $a$ to
$B$, and $B$ sends packet $b$ to $A$; but both $a$ and $b$ must
first be sent to $R$, which then forwards each packet. Without COPE,
the relay $R$ must send each packet individually using two time slots.
With COPE, the relay $R$ will generate one coded message, $a$$\oplus$$b$
(where $\oplus$ indicates mod 2 addition), and will broadcast this
packet to both $A$ and $B$ in a single time slot. Since both nodes
have their original packets, they can both decode the message and
extract $b$ and $a$ respectively. When we consider broadcast traffic
and some nodes requiring multiple degrees of freedom, we generalize
COPE by allowing for a larger field size in order to transmit different
linear combinations of packets. When considering the use of MPR, we
allow both $A$ and $B$ to send their respective packets to $R$
in the same time slot. $R$ can then code the two packets together
and transmit a single coded message back to $A$ and $B$. In the
remainder of the paper, we apply this concept to provide an analysis
of several 5-node topology components.}

\begin{figure}
\begin{centering}
\textcolor{black}{\includegraphics[width=0.7\columnwidth]{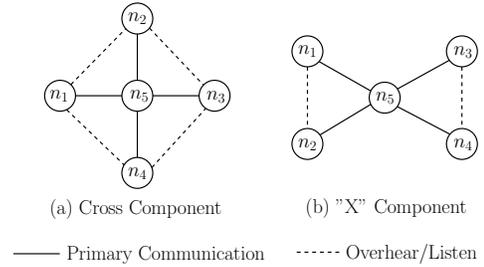}}
\par\end{centering}

\textcolor{black}{\caption{\label{fig:Network-Topologies}\textcolor{black}{Basic network structures
responsible for traffic bottlenecks and congestion in larger networks.
We analyze these components and variants of them.}}
}
\end{figure}

\textcolor{black}{We use the basic network topology components shown
in Fig. \ref{fig:Network-Topologies} since these are the primary
network structures in large networks that form bottlenecks and create
congestion. }We first analyze the throughput behavior of using network
coding and MPR using these small 5-node components and then generalize
for components with $N$ nodes, shown in Fig. \ref{fig:Generalized-topology-components},
so that we can analyze the throughput behavior in the asymptotic regime.\textcolor{black}{{}
We focus our attention on traffic that travels through the center
node so that we model both bottlenecks and multi-hop networks. Within
our model, each node randomly generates a packet and then transmits
it through the relay node to its destination. The relay is fully connected
regardless of the topology, and packets generated at the relay require
only a single hop to reach their destination within the topology component.
Each topology component has specific constraints due to their structure.
In Fig. \ref{fig:Network-Topologies}, we define these constraints
through the use of a solid edge that depicts active, or primary communication,
and a dotted edge that depicts passive, or overhear/listening communication.
The absence of an edge between any two nodes indicates that all communication
between the two nodes must be routed through a relay. Within the cross
topology component, each traffic flow originating from a given node
is terminated at the node directly opposite the center; and in the
``X'' topology component, all flows originating from a node in a given
set terminates at a node in the opposite set. Therefore, each flow
must pass through the center regardless of topology. For example,
nodes $n_{1}$, $n_{2}$, and $n_{5}$ in the {}``X'' topology component
are fully connected and nodes $n_{3}$, $n_{4}$, and $n_{5}$ are
also fully connected; but $n_{1}$ and $n_{2}$ are not connected
to $n_{3}$ and $n_{4}$. All traffic between any node $\{n_{1},n_{2}\}\in X_{1}$
and a node $\{n_{3},n_{4}\}\in X_{2}$ must travel through the center.}%

\textcolor{black}{We model the MAC using the primary characteristics
of the 802.11 MAC based on the empirical results from \citep{Katti00}.
The non-monotonic behavior in the experimental throughput shown in
the upper half of Fig. \ref{fig:Emperical-COPE-performance} is a
result of both collisions and fairness imposed by the 802.11 MAC.
Since the effects of collisions on throughput are small in relation
to the effects of the 802.11 MAC fairness mechanisms, we do not consider
collisions due to either hidden nodes or identical back off times.
As a result, the model inherently underestimates the full benefits
of implementing MPR, which reduces collisions, in wireless networks.
Furthermore, we do not consider the overhead associated with the virtual
802.11 CS mechanisms (RTS/CTS) when analyzing unicast traffic. The
following analysis of the throughput gains produced through the use
of MPR and network coding is a lower-bound to the achievable gain
as a result of these model assumptions. }

\textcolor{black}{The current 802.11 MAC protocol's goal is to distribute
time slots equally among all competing }\textit{\textcolor{black}{nodes}}\textcolor{black}{{}
within a network, regardless of topology, and does not consider fairness
of information }\textit{\textcolor{black}{flows}}\textcolor{black}{.
As network load increases, the MAC limits each edge node's traffic
to the center, or relay, while the rate of traffic from sources directly
connected to the center (self-generated traffic) will not be similarly
constrained. Nodes sending both relayed traffic and self-generated
traffic in each topology component will therefore inherently send
more of their own self-generated traffic, and the effectiveness of
network coding will be reduced. This type of allocation occurs in
the first part of our analysis, and we propose a modified MAC approach
in Section \ref{sec:Improving-the-MAC} that improves throughput by
allocating time slots proportional to information flows.}

\textcolor{black}{Since our main focus is the interaction between
network coding and MPR, we will not address specific implementation
methods for MPR. We assume that each node receives multiple simultaneous
packets without delay or loss. It is further assumed that the wireless
channel is lossless, feedback is perfect, and the load required for
acknowledgments is contained as part of the initial transmission's
load. If a node is not transmitting and it has primary communication
or can overhear another node, it will automatically overhear any transmission
made by that node and use the information to decode any coded messages
it receives. In addition, packet transmission is never delayed. If
a node does not have more than one codable packet, it does not wait
for another codable packet to arrive. Rather, it sends the packet
uncoded at the first opportunity. Finally, all packets headed towards
the same next hop will not be coded together because the next hop
would not be able to decode such coded packets.}

\textcolor{black}{The total offered load $P$ to the network from
the set of source nodes $i\in N$, is defined as $P=\sum_{i\in N}\rho_{i}$,
where $N$ is the total number of nodes in the topology component
and $\rho_{i}=\nicefrac{k_{i}}{100}$ is each node's individual load
contribution to the network, or the fraction of time required to send
all of its $k_{i}$ packets to the next hop. We stochastically determine
$k_{i}$ using a binomial distribution given $P$ in each iteration
of our simulation and average these results for each total offered
load evaluated. }

\textcolor{black}{We assume each node transmits all of its packets
to the center node $n_{5}$. Once every node ($i\neq5$) has sent
all of its packets, the center node will either identify coding opportunities
and transmit a set of coded messages optimized for the topology component
or send the packets uncoded. When MPR is used, we allow $m$ packets
to be sent from different sources in a single time slot. Since MPR
provides a method of avoiding collisions due to hidden nodes, we use
the existing CSMA/CA protocols employed by 802.11 for each $m=2$
case. Cases involving $m=4$ requires an extension to CSMA/CA to allow
each edge node to transmit in the same time slot to $n_{5}$. In addition,
the results shown in the figures found in subsequent sections are
averaged over the packet arrival distribution and do not reflect the
maximum achievable gains that occur when $k_{i}=k_{j}$ $\forall i,j$,
$i\neq j$. }

\textcolor{black}{We also consider a unicast transmission complete
when all packets from each source node successfully reach their destinations;
and broadcast transmissions complete when all nodes have received
each packet from all sources. Furthermore, each node is half-duplex,
and as a result, a node cannot receive other node's transmissions
while it is transmitting.}

\section{\textcolor{black}{Multi-Packet Reception and Network Coding Performance
Analysis\label{sec:Multi-Packet-Reception-and}}}

\textcolor{black}{With each of the network topology components shown
in Fig. \ref{fig:Network-Topologies}, we analyze the topology component
performance with and without the use of network coding and MPR. We
also consider both unicast and broadcast traffic.}

\subsection{\textcolor{black}{Cross Topology Component Analysis\label{sub:Cross-Network-Topology}}}

\textcolor{black}{Each node $i\in[1,5]$, requires $\rho_{i}$ of
the time to send all of its packets one hop where $\rho_{i}$ is the
initial load originating from node $i$. The center node $n_{5}$
requires $\rho_{5}$ of the time to send its own packets one hop plus
the load $\rho_{R}$ required to route all traffic from the edge nodes
to their final destination within the topology component. Let $P=\sum_{i=1}^{5}\rho_{i}$
where $\rho_{i}$ is stochastically determined according to the binomial
distribution described in Section \ref{sec:Network-models}; let the
relay load be $\rho_{R}=\frac{1}{c}\sum_{j=1}^{4}\rho_{j}$ for $j\in[1,4]$
where $c$ is the number of packets that can be effectively coded
together; and let the total network component load $P_{T}$ required
to send all packets to their intended destinations be $P_{T}=P+\rho_{R}$.
In the case of the cross topology component and enough packets to
code together: $c=4$ for $m=1$; $c=4$ for $m=2$ when opposite
nodes transmit at the same time (i.e., CSMA is used); $c=2$ for $m=2$
when no restrictions are placed on the order of transmission from
each node (i.e., CSMA is not used); and $c=2$ for all $m\geq3$.
When enough codable packets do not exist, the coding coefficient $c$
will equal the maximum number of packets that can be coded together.
Let the fraction of allocated time slots a node receives as a result
of the MAC be $s_{i}$.}

\begin{figure}
\begin{centering}
\textcolor{black}{\includegraphics[width=3in]{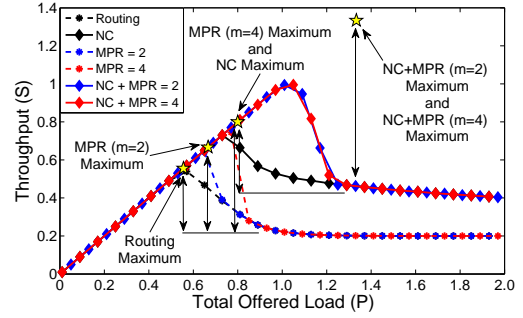}}
\par\end{centering}

\textcolor{black}{\caption{\label{fig:Cross-OriginalMAC}Average unicast and broadcast throughput
for a 5-node cross topology component. Each vertical double arrow
shows the difference in the maximum and saturated throughput due to
MAC fairness for each case.}
}
\end{figure}

\textcolor{black}{The throughput $S$ for the cross topology component
depicted in Fig. \ref{fig:Network-Topologies}(a) with unicast and
broadcast traffic is shown as a function of $P$ in Fig. \ref{fig:Cross-OriginalMAC}.
The throughput shown in this figure is averaged over the loads obtained
using the distribution discussed in Section \ref{sec:Network-models};
the stars depict the maximum achievable throughput when the MPR and/or
network coding gain is maximized. When $P_{T}<1$, each node is allocated
enough time slots to send all of its packets, and the allocated load
is $s_{j}=\rho_{j}$ for $j\in[1,4]$ and $s_{5}=\rho_{5}+\rho_{R}$.
The throughput $S$ increases linearly as the network load increases,
regardless of the use of MPR or network coding. The throughput for
each case reaches a maximum when $P_{T}=1$ and transitions into a
saturated region for $P_{T}>1$, where for each node, the allocated
load $s_{j}\leq\rho_{j}$ and $s_{5}\leq\rho_{5}+\rho_{R}$. When
network coding is not used, the throughput is $S=s_{5}$; and when
network coding is used, the throughput will be a function of the number
of packets that can be effectively coded together.}%

\subsubsection{\textcolor{black}{Routing (No Network Coding, $m=1$)}}

\textcolor{black}{We will use routing as the baseline for our analysis.
Consistent with the results found in \citep{Katti00} and the analysis
performed in \citep{Zhao00}, the throughput increases linearly within
the non-saturated region, $P\in[0,\nicefrac{5}{9})$. At $P=\nicefrac{5}{9}$,
the throughput is maximized. For example, consider the situation where
the source loads are symmetric, i.e., each node has an equal number
of packets to send. The maximum throughput of $S=\nicefrac{5}{9}$,
depicted by a star in Fig. \ref{fig:Cross-OriginalMAC}, occurs when
each source reaches $\rho_{i}=\nicefrac{1}{9}$ for $i\in[1,5]$.
The total load of the center node, as a consequence, is $\rho_{5}+\rho_{R}$
where $\rho_{R}=\sum_{i=1}^{4}\rho_{j}=\nicefrac{4}{9}$ for $j\in[1,4]$.
Since $P_{T}=1$, $s_{j}=\rho_{j}$ and $s_{5}=\rho_{5}+\rho_{R}$.}

\textcolor{black}{The throughput saturates for $P>\nicefrac{5}{9}$.
Initially, the 802.11 MAC allocates time slots to nodes requiring
more resources. The throughput is therefore the amount of time $n_{5}$
is able to transmit, $s_{5}=1-\sum_{i=1}^{4}s_{i}$, which decreases
as $P$ increases. The network component completely saturates when
each node requires a large fraction of the available time slots. The
MAC restricts each node's access to the channel by ensuring fairness
among all nodes, i.e., $s_{i}=\nicefrac{1}{5}$ for $i\in[1,5]$.
The total saturated throughput is equal to the total amount of information
that $n_{5}$ transmits, i.e., $S=s_{5}=\nicefrac{1}{5}$.}

\subsubsection{\textcolor{black}{Network Coding Only ($m=1$)}}

\textcolor{black}{We now allow network coding to be used by the center
node. Each edge node transmits one at a time to the center node, allowing
the two nodes within range of the transmitting node to use opportunistic
listening to overhear and store each transmitted packet. After each
edge node has completed transmission, $n_{5}$ transmits a single
coded packet which is sufficient for each edge node to obtain the
single degree of freedom it still requires.}

\textcolor{black}{From Fig. \ref{fig:Cross-OriginalMAC}, when $P\in[0,\nicefrac{5}{9})$,
network coding is seen to provide no additional gains over the use
of routing alone since $n_{5}$ can forward each packet received without
the MAC limiting its channel use. For $P\in[\nicefrac{5}{9},\nicefrac{5}{6}),$
network coding is instrumental in achieving the throughput shown.
The MAC does not limit channel resources until the maximum throughput
of $S=\nicefrac{5}{6}$ is reached when $P_{T}=\sum_{i=1}^{5}\rho_{i}+\frac{1}{4}\sum_{j=1}^{4}\rho_{j}=1$.
At this maximum, the MAC ensures fairness among all competing nodes
and the throughput saturates.}

\textcolor{black}{As both $P$ and $\rho_{5}$ increase, the gain
provided by network coding diminishes. The number of packets reaching
$n_{5}$ from each edge node is limited by the MAC while packets introduced
into the network component by $n_{5}$ are not. The coding gain, therefore,
approaches zero as $P\rightarrow\infty$.}

\subsubsection{\textcolor{black}{Multi-Packet Reception of Order 2 and 4 (No Network
Coding and $m=2,4$)}}

\textcolor{black}{MPR is similar to the routing case described earlier
except we now allow a maximum of $m$ edge nodes to transmit within
a given time slot. For $m=2$, the total time used by all of the edge
nodes to transmit their packets to $n_{5}$ is $\nicefrac{1}{2}$
that needed by routing while the center node cannot transmit multiple
packets simultaneously and must transmit each received packet individually.
Using CSMA, which restricts nodes opposite each other to transmit
at the same time, the point at which the protocol saturates for symmetric
source loads occurs when $\rho_{i}=\nicefrac{1}{7}$ for $i\in[1,5]$
and $\rho_{R}=\sum_{j=1}^{4}\rho_{j}=\nicefrac{4}{7}$. This maximum,
which yields a throughput of $S=\nicefrac{5}{7}$, occurs when each
source has equal loads and is not reflected in Fig. \ref{fig:Cross-OriginalMAC}
because the throughput shown is averaged over the packet arrival distribution
explained in Section \ref{sec:Network-models}. The throughput saturates
to the same throughput as routing for values of $P_{T}>1$ and the
gain for $m=2$ is 1 due to the suboptimal saturation behavior of
the protocol.}

\textcolor{black}{The behavior for $m=4$ is the same as that for
$m=2$ except the maximum of $S=\nicefrac{5}{6}$ occurs when $\rho_{i}=\nicefrac{1}{6}$
and $\rho_{R}=\sum_{j=1}^{4}\rho_{j}=\nicefrac{2}{3}$. We allow all
of the edge nodes to transmit their packets to $n_{5}$ simultaneously,
requiring a total of $\nicefrac{1}{6}$ of the time slots. Node $n_{5}$
then sends each node's packet individually, including its own, to
the intended recipient requiring the remainder of the time slots to
finish each unicast/broadcast transmission. As $P$ increases, the
MAC limits each node's number of available time slots and $S$ saturates
to $\nicefrac{1}{5}$. Again, the gain in the saturated region for
$m=4$ is equal to the cases of $m=2$ and routing.}

\textcolor{black}{The gain as a result of the use of MPR depends on
an adequate number of source nodes with information to send. If $m$
is greater than the total number of nodes with information to send,
i.e., $m>N$, the MPR gain will be less than when $m\leq N$. In addition,
the achievable gain for implementations using stochastic message arrival
and transmission times will be upper-bounded by the results shown
in this section and lower-bounded by the throughput for the non-MPR
(routing) case seen in Fig. \ref{fig:Cross-OriginalMAC}.}

\subsubsection{\textcolor{black}{Network Coding with Multi-Packet Reception of Order
2 and 4 ($m=2,4$)}}

\textcolor{black}{The case when MPR is combined with network coding
results in further improvement as seen in Fig. \ref{fig:Cross-OriginalMAC}.
Unlike the case where we considered MPR alone, the order in which
each node transmits and symmetric traffic across the topology component
is crucial to achieving the maximum throughput gain. As a result,
we continue to use CSMA to ensure nodes opposite the center transmit
at the same time so that we both facilitate opportunistic listening
and enable coding opportunities by $n_{5}$. The average throughput
shown in Fig. \ref{fig:Cross-OriginalMAC} for both cases discussed
in this section do not reach the maxima found below and indicated
by a star in the figure because of the stochastic load distribution,
which results in asymmetric traffic among the set of nodes. Should
each node have an equal amount of information to send, the maxima
found below will be reached.}

\textcolor{black}{When $m=2$, the maximum throughput of $S=\nicefrac{5}{4}$
occurs when $\rho_{i}=\nicefrac{1}{4}$ for $i\in[1,5]$ and $\rho_{R}=\nicefrac{1}{4}\sum_{j=1}^{4}\rho_{j}=\nicefrac{1}{4}$.
Each set of nodes, $\{n_{1},n_{3}\}$ and $\{n_{2},n_{4}\}$, uses
$\nicefrac{1}{4}$ of the total number of time slots to transmit to
$n_{5}$ which then transmits a single coded packet derived from all
four node's native packets plus its own packets. For $P_{T}>1$, the
throughput saturates to the saturated network coding throughput due
to the MAC. The saturated gain for $m=2$ is therefore equal to the
gain found when network coding was used alone in this region.}

\textcolor{black}{The throughput using network coding and $m=4$ for
unicast traffic is equivalent to network coding and $m=2$. This throughput
can be achieved using one of two methods. We force all four edge nodes
to transmit to $n_{5}$ which then transmits two coded packets in
addition to its own; or we limit the number of simultaneous transmissions
to two thus allowing $n_{5}$ to code everything together and send
a single coded packet to all of the edge nodes. Either strategy will
achieve the same throughput gain although the difference occurs when
considering either unicast (former option) or broadcast (later option).
The maximum throughput for broadcast traffic using the first method
is $S=1$ and $S=\nicefrac{5}{4}$ for the second which is consistent
with the maximum unicast throughput.}

\subsection{\textcolor{black}{{}``X'' Topology Component}}

\textcolor{black}{}%
\begin{figure}
\centering{}\textcolor{black}{\includegraphics[width=3in]{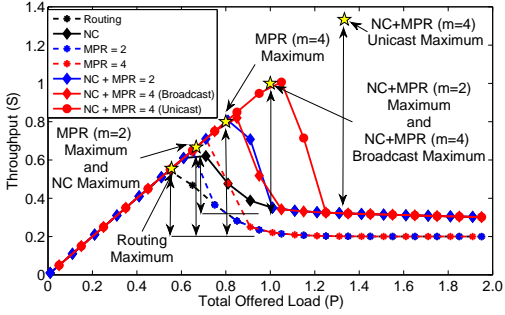}\caption{\label{fig:X-OriginalMAC}Average broadcast and unicast throughput
for a 5-node {}``X'' topology component. Each vertical double arrow
shows the difference in the maximum and saturated throughput due to
MAC fairness for each case.}
}
\end{figure}
\textcolor{black}{The throughput for the {}``X'' topology component,
Fig. \ref{fig:Network-Topologies}(b), is shown in Fig. \ref{fig:X-OriginalMAC}.
It can be easily verified that the routing and $m=2$ and $4$ cases
for this topology component are the same as the cross topology component.
We focus on the cases incorporating only network coding.}

\subsubsection{\textcolor{black}{Network Coding Only ($m=1$)}}

\textcolor{black}{Limiting the ability to overhear other edge nodes
in the topology component results in a reduction of the number of
native packets that can be coded together. Packets from different
nodes within the same set, i.e., $\{n_{1},n_{2}\}\in X_{1}$ and $\{n_{3},n_{4}\}\in X_{2}$,
cannot be coded together therefore restricting $n_{5}$ from coding
all of the edge node packets together. The center node must make a
minimum of two transmissions for every 4 packets it receives from
different edge nodes in order to ensure that each destination node
has the required degrees of freedom to decode the appropriate packets.}

\textcolor{black}{The throughput of the {}``X'' topology component
increases linearly until it reaches its maximum at $S=\nicefrac{5}{7}$.
Assuming symmetric source loads, this maximum occurs when $\rho_{i}=\nicefrac{1}{7}$
for $i\in[1,5]$ and $\rho_{R}=\nicefrac{1}{2}\sum_{j=1}^{4}\rho_{i}=\nicefrac{2}{7}$.
The throughput saturates for $P_{T}>1$ and the non-monotonic behavior
is again due to the fairness aspect of the 802.11 MAC. Using this
topology component, it is evident that the protocols employed by 802.11
systems restrict the total throughput when the network is saturated
and gains can be achieved by modification of the existing MAC.}

\subsubsection{\textcolor{black}{Network Coding with Multi-Packet Reception of Order
2 and 4 ($m=2,4)$}}

\textcolor{black}{The throughput for the case in which network coding
is used in conjunction with MPR ($m=2$) for the {}``X'' topology
component is similar to the cross topology component throughput. Using
CSMA, the throughput increases linearly until it reaches its maximum
at $S=1$ when $\rho_{i}=\nicefrac{1}{5}$ for $i\in[1,5]$ and $\rho_{R}=\nicefrac{1}{2}\sum_{j=1}^{4}\rho_{j}=\nicefrac{2}{5}$.
The throughput for this case saturates to the network coding throughput
for $P_{T}>1$.}

\textcolor{black}{The average and maximum throughput shown in Fig.
\ref{fig:X-OriginalMAC} for $m=2$ is achieved for both unicast and
broadcast traffic when using CSMA to force nodes from different sets
to transmit to $n_{5}$ at the same time. Removing this constraint
results in the same throughput for unicast traffic. Broadcast traffic
throughput will be upper bounded by the unicast throughput and lower
bounded by the $m=2$ without network coding case. Furthermore, the
broadcast throughput will be dependent on the mechanism of determining
the order of transmissions, such as CSMA, round-robin, or other similar
scheme, within the wireless channel.}

\textcolor{black}{For $m=4$, the maximum unicast throughput of $S=\nicefrac{5}{4}$
is achieved when allowing all four source nodes to transmit to the
center at the same time. The center node codes a maximum of two native
packets together from different source node sets and transmits two
coded packets back to the edge nodes, including its own uncoded packets,
in order to complete the unicast transmission. At the completion of
the unicast transmission, each node still requires a maximum of one
additional degree of freedom to complete the broadcast transmission.
Allowing $n_{5}$ to code all of the native edge node packets together
and send one additional coded transmission enables each node to extract
the required degree of freedom and obtain the full set of transmitted
messages. The maximum throughput for this case is therefore the same
as the case for network coding with $m=2$ and is equal to $S=1$.
Similar to the cross topology component, the average throughput for
both cases discussed in this section does not reach the maxima found
because of the stochastic load distribution, which results in asymmetric
traffic flows across the center node. If the each node had an equal
amount of information to send, then the maxima found in this section
would be achieved.}

\begin{figure}
\begin{centering}
\textcolor{black}{\includegraphics[width=3in]{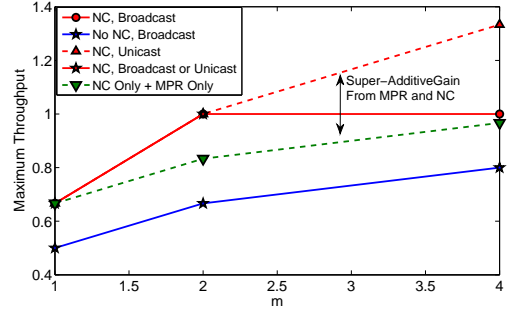}}
\par\end{centering}

\textcolor{black}{\caption{\textcolor{black}{Maximum throughput of a 5-node {}``X'' topology
component as function of the MPR capability. Super-additive gains
are achieved when using network coding in conjunction with MPR.\label{fig:sat-flow}}}
}
\end{figure}

\textcolor{black}{Fig. \ref{fig:sat-flow} shows a summary of our
analysis by plotting the maximum unicast and broadcast throughput
as a function of the MPR capability.. In addition, it illustrates
the super-additive behavior of the throughput when MPR is used in
conjunction with network coding by comparing this throughput with
the throughput that would be obtained by adding the individual gains
obtained using MPR and network coding separately.}%

\subsection{\textcolor{black}{Partial Topology Components}}

The removal of an overhear/listen edge in both topology components
found in Fig. \ref{fig:Network-Topologies} has little impact on the
throughput gain. In the case of the cross topology component, the
removal of a single edge results in a maximum throughput found using
the unmodified {}``X'' topology component. In the case of the {}``X''
topology component, the gain resulting from the use of network coding
is reduced; and as a result, the throughput decreases. It can be verified
that the \textcolor{black}{the maximum throughput for the case where
network coding and $m=2$ is $S=1$ for unicast traffic and $S=\nicefrac{5}{6}$
for broadcast traffic. This is only a slight reduction in throughput
from the unmodified {}``X'' topology component's throughput. On
the other hand when $m=4$, the maximum is the same as that found
for the partial cross and {}``X'' topology components. Since MPR
restricts each node's ability to overhear other node's transmissions,
the limitations imposed by the network topology do not impact the
gain provided by the combined use of MPR ($m=4$) and network coding.}

\section{\textcolor{black}{Improving the MAC Fairness Protocol\label{sec:Improving-the-MAC}}}

\textcolor{black}{While several approaches to improve fairness among
flows in 802.11 networks have been suggested, none have considered
the combined use of MPR and network coding. As a result, our approach
optimizes the throughput of these networks subject to MPR, network
coding, the topology component, and fairness to }\textit{\textcolor{black}{flows}}\textcolor{black}{{}
rather than to }\textit{\textcolor{black}{nodes}}\textcolor{black}{.
The basic premise behind the improved protocol approach is to allocate
resources proportional to the amount of non-self-generated traffic
flowing through each node when the network saturates. While allocating
fewer resources to flows originating at the center and more resources
to flows originated at edge nodes yields even higher throughput, our
policy ensures that each flow of information is given the same priority.
The center node will be allocated more resources than each edge node
in order to relay information; but it must also limit the amount of
self-generated traffic so that it equals the average per node non-self-generated
traffic being relayed. }

\begin{figure}
\begin{centering}
\includegraphics[width=0.7\columnwidth]{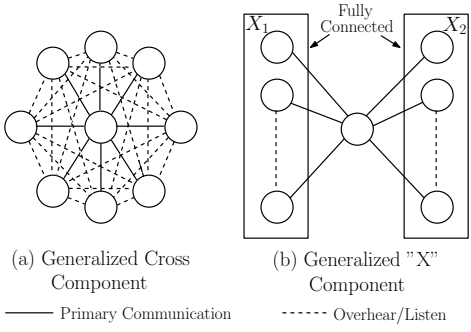}
\par\end{centering}

\textcolor{black}{\caption{\textcolor{black}{\label{fig:Generalized-topology-components}Generalized
topology components for $N$ nodes.}}
}

\end{figure}

\textcolor{black}{We design the revised MAC using a slight modification
of the components found in Fig. \ref{fig:Network-Topologies}. For
the cross topology component, we let there be $N-1$ edge nodes and
a single center, or relay, node. All edge nodes are connected with
the center node and connected with all other edge nodes except the
one directly opposite the center. For the {}``X'' topology component,
we also let there be $N-1$ edge nodes and a single center node. The
edge nodes are split into two sets $X_{1}$ and $X_{2}$. All edge
nodes within a given set are fully connected and also connected to
the center node. The MAC is optimized for traffic that travels through
the center node. Within the cross topology component, each node communicates
with the node directly opposite the center. In the {}``X'' topology
component, each node communicates with a node in a different set.
Fig. \ref{fig:Generalized-topology-components} provides an illustration
of both generalized components.}%

\textcolor{black}{We define the throughput $S$, which is analogous
to the throughput defined in Section \ref{sec:Multi-Packet-Reception-and},
as the total number of nodes transmitting data $N$ divided by the
total number of time slots needed to complete either all unicast or
broadcast sessions:\begin{equation}
S=\dfrac{N}{N_{MPR}+N_{C}}\label{eq:Throughput}\end{equation}
The denominator $N_{MPR}+N_{C}$ is just the sum of the number of
time slots required from the $N-1$ edge nodes to the center plus
the number of time slots required from the center to the edge nodes.
The former is determined by the MPR coefficient and the later is determined
by network coding. The number $N_{MPR}$ of time slots required from
the edge nodes to the center is dependent on the structure of the
topology component and the implementation of the MAC. The number $N_{C}$
of time slots required from the center is the maximum degrees of freedom
that any given node requires in order to decode each coded packet.
With the cross topology component with network coding, the term $N_{C}=1+(m-1)=m$
where the first term in the sum is a result of the flow originating
at the center node, and the second term $(m-1)$ comes from the fact
that each edge node was able to overhear all but $m-1$ degrees of
freedom from the rest of the edge nodes. With the {}``X'' topology
component with network coding, the term $N_{C}=\max\left(\left|X_{1}\right|,\left|X_{2}\right|\right)+1$
where the first term is the cardinality of the maximum set of edge
nodes representing the maximum degrees of freedom that the center
must send to the edge nodes and the second term results from the flow
originating at the center node.}

\begin{figure}
\centering{}\textcolor{black}{\includegraphics[width=3.5in]{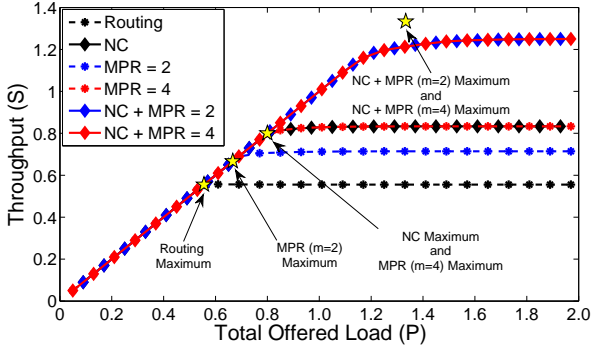}\caption{\label{fig:5Node-Cross_ImprovedMAC}5-Node cross topology component
unicast and broadcast throughput using the improved MAC}
}
\end{figure}

\begin{figure}
\begin{centering}
\textcolor{black}{\includegraphics[width=3.5in]{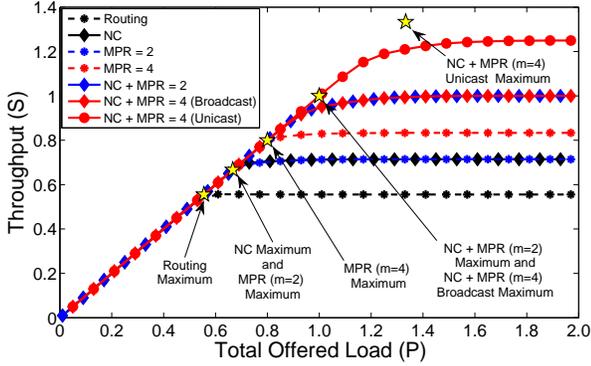}}
\par\end{centering}

\textcolor{black}{\caption{\label{fig:5-Node-X-Unicast}5-Node {}``X'' component throughput
using the improved MAC}
}
\end{figure}

\textcolor{black}{The allocated number of time slots each node receives
so that the throughput $S$ is maximized, subject to the flow constraints
and $\sum_{j=1}^{N-1}\nicefrac{s_{j}}{m}+s_{R}=1$, is divided into
three cases where where $s_{j}$ is the fraction of time slots allocated
to each edge node and $s_{R}$ is the fraction of time slots allocated
to the center node. Similar to Section \ref{sec:Multi-Packet-Reception-and},
the throughput $S=s_{R}$ when network coding is not used, and $S$
is a function of the number of packets that can be effectively coded
together, which is dependent on the MPR coefficient $m$, the use
of CSMA, and the traffic type (unicast or broadcast), when network
coding is used. The cases include:}
\begin{itemize}
\item \textsl{\textcolor{black}{Cross Topology Component with Unicast Traffic
or Broadcast Traffic:}}\textcolor{black}{{} Assuming that there are
no constraints on the order in which each node transmits to the center
node, the allocation of resources is the same for both unicast and
broadcast sessions. Without network coding, the center node will require
a number of time slots equal to the number of source nodes $N$. With
network coding, throughput is maximized by ensuring the center node
codes the maximum number of native packets together. Implementation
of MPR can potentially prevent each node from immediately decoding
any coded message sent by the center since we are allowing nodes with
the ability to overhear each other the ability to transmit at the
same time. When $m=2$, the center node needs to send two coded packets,
each combined in a different manner, to ensure that each edge node
has the necessary degrees of freedom to decode each packet. Generalizing
for $N$ and $m$ as well as considering only integer numbers of time
slots:\begin{equation}
s_{j}=\begin{cases}
\frac{1}{\lceil(N-1)/m\rceil+N} & \textrm{without NC}\\
\frac{1}{\lceil(N-1)/m\rceil+m_{c}+1} & \textrm{with NC}\end{cases}\label{eq:cross_bw1}\end{equation}
and\begin{equation}
s_{R}\leq\begin{cases}
\frac{N}{\lceil(N-1)/m\rceil+N} & \textrm{without NC}\\
\frac{m_{c}+1}{\lceil(N-1)/m\rceil+m_{c}+1} & \textrm{with NC}\end{cases}\label{eq:cross_bw2}\end{equation}
We define $m_{c}=m$ for $m=1$ and $m=2$ when CSMA is not used and
$m_{c}=m-1$ for $m=2$ when CSMA is used so that only nodes opposite
the center are allowed to transmit in the same time slot. In addition,
the term $m_{c}=m-1$ for all situations where $m=4$. Furthermore,
(\ref{eq:cross_bw2}) is met with equality if CSMA is used for $m=1$
and 2 as well as for all cases when $m=4$. Equation (\ref{eq:cross_bw2})
may be met with inequality when CSMA is not used for $m=2$ since
there is a non-zero probability that any given node may miss a packet
from a node in which it can overhear while it is transmitting. Using
a scheme such as CSMA results in a significant throughput gain for
small $N$ but becomes insignificant as $N$ grows.}
\item \textsl{\textcolor{black}{{}``X'' Topology Component: }}\textcolor{black}{The
fraction of time slots $s^{U}$ allocated to each node for unicast
traffic is:\begin{equation}
s_{j}=s_{j}^{U}=\begin{cases}
\frac{1}{\lceil(N-1)/m\rceil+N} & \textrm{without NC}\\
\frac{1}{\lceil(N-1)/m\rceil+\max\left(\mid X_{1}\mid,\mid X_{2}\mid\right)+1} & \textrm{with NC}\end{cases}\label{eq:x_bw1}\end{equation}
and \begin{equation}
s_{R}=s_{R}^{U}=\begin{cases}
\frac{N}{\lceil(N-1)/m\rceil+N} & \textrm{without NC}\\
\frac{\max\left(\mid X_{1}\mid,\mid X_{2}\mid\right)+1}{\lceil(N-1)/m\rceil+\max\left(\mid X_{1}\mid,\mid X_{2}\mid\right)+1} & \textrm{with NC}\end{cases}\label{eq:x_bw2}\end{equation}
When considering broadcast traffic, additional degrees of freedom
must be sent by the center to complete the session. Without network
coding, equations \ref{eq:x_bw1} and \ref{eq:x_bw2} still hold.
With network coding, there is a possibility that each destination
node will require a maximum of one additional degree of freedom per
node for $m=2$ or three degrees of freedom per node for $m=4$ when
either $\mid X_{1}\mid\geq m$ or $\mid X_{2}\mid\geq m$ and the
order of node transmission is not enforced (i.e., CSMA is not used).
Providing these additional degrees of freedom can be accomplished
by the center node sending at most three additional coded packets,
where each coded packet contains a different combination of all of
the native edge node packets. Each edge node's fraction of time slots
is maximized when the cardinality of each set, $X_{1}$ and $X_{2}$,
are equal, and minimized when the cardinalities differ most and transmission
from the edge nodes to the center is asymmetric, i.e., multiple nodes
from a single set transmit at the same time. The fraction of time
slots each node receives for broadcast traffic, $s^{B}$, with network
coding is then bounded by: \begin{equation}
{\textstyle s_{j}^{U}\geq s_{j}^{B}\geq\frac{1}{\lceil(N-1)/m\rceil+\max\left(\mid X_{1}\mid,\mid X_{2}\mid\right)+m}}\end{equation}
and \begin{equation}
s_{R}^{U}{\textstyle \leq s_{R}^{B}\leq\frac{\max\left(\mid X_{1}\mid,\mid X_{2}\mid\right)+m}{\lceil(N-1)/m\rceil+\max\left(\mid X_{1}\mid,\mid X_{2}\mid\right)+m}}\end{equation}
}
\end{itemize}
\textcolor{black}{We applied our revised fairness protocol to both
the 5-node cross and {}``X'' topology components using the same
model described in Section \ref{sec:Multi-Packet-Reception-and}.
In addition, the throughput $S$ can be calculated using the methods
described in Section \ref{sec:Multi-Packet-Reception-and} and earlier
in this section for both the non-network coding and network coding
cases. We find that the throughput saturates at the }\textit{\textcolor{black}{maxima}}\textcolor{black}{{}
found in Section \ref{sec:Multi-Packet-Reception-and} for each topology
component. Fig. \ref{fig:5Node-Cross_ImprovedMAC} and \ref{fig:5-Node-X-Unicast}
show both the unicast and broadcast throughput for the cross and {}``X''
topology components, respectively, using our improved MAC approach.
The gains associated with our modification of the fairness protocol
are listed in Table \ref{tab:X-MAC-Gain}.}

\textcolor{black}{}%
\textcolor{black}{}%
\textcolor{black}{}%
\begin{table}
\begin{centering}
\textcolor{black}{}\begin{tabular}{|c|c|c|c|c|}
\hline 
\textcolor{black}{\footnotesize Case} & \multicolumn{2}{c|}{\textcolor{black}{\footnotesize (a) Cross Component}} & \multicolumn{2}{c|}{\textcolor{black}{\footnotesize (b) {}``X'' Component}}\tabularnewline
\cline{2-5} 
 & \textcolor{black}{\footnotesize Unicast} & \textcolor{black}{\footnotesize Broadcast} & \textcolor{black}{\footnotesize Unicast} & \textcolor{black}{\footnotesize Broadcast}\tabularnewline
\hline
\hline 
\textcolor{black}{\footnotesize Routing} & \textcolor{black}{\footnotesize 2.8} & \textcolor{black}{\footnotesize 2.8} & \textcolor{black}{\footnotesize 2.8} & \textcolor{black}{\footnotesize 2.8}\tabularnewline
\hline 
\textcolor{black}{\footnotesize Network} & \textcolor{black}{\footnotesize 4.2} & \textcolor{black}{\footnotesize 4.2} & \textcolor{black}{\footnotesize 3.6} & \textcolor{black}{\footnotesize 3.6}\tabularnewline
\textcolor{black}{\footnotesize Coding (NC)} &  &  &  & \tabularnewline
\hline 
\textcolor{black}{\footnotesize MPR } & \textcolor{black}{\footnotesize 3.6} & \textcolor{black}{\footnotesize 3.6} & \textcolor{black}{\footnotesize 3.6} & \textcolor{black}{\footnotesize 3.6}\tabularnewline
\textcolor{black}{\footnotesize $(m=2)$} &  &  &  & \tabularnewline
\hline 
\textcolor{black}{\footnotesize MPR } & \textcolor{black}{\footnotesize 4.2} & \textcolor{black}{\footnotesize 4.2} & \textcolor{black}{\footnotesize 4.2} & \textcolor{black}{\footnotesize 4.2}\tabularnewline
\textcolor{black}{\footnotesize $(m=4)$} &  &  &  & \tabularnewline
\hline 
\textcolor{black}{\footnotesize NC and MPR} & \textcolor{black}{\footnotesize 6.3} & \textcolor{black}{\footnotesize 6.3} & \textcolor{black}{\footnotesize 5.0} & \textcolor{black}{\footnotesize 5.0}\tabularnewline
\textcolor{black}{\footnotesize{} $(m=2)$} &  &  &  & \tabularnewline
\hline 
\textcolor{black}{\footnotesize NC and MPR} & \textcolor{black}{\footnotesize 6.3} & \textcolor{black}{\footnotesize 6.3} & \textcolor{black}{\footnotesize 6.3} & \textcolor{black}{\footnotesize 5.0}\tabularnewline
\textcolor{black}{\footnotesize{} $(m=4)$} &  &  &  & \tabularnewline
\hline
\end{tabular}
\par\end{centering}

\textcolor{black}{\caption{\label{tab:X-MAC-Gain}Gains in the saturated network throughput with
the improved MAC. Each gain is baselined against the saturated routing
throughput using the original 802.11 MAC.}
}
\end{table}

\section{\textcolor{black}{Performance of Network Coding and MPR with Large
N\label{sec:Performance-of-large-N}}}

\textcolor{black}{The gain provided by using MPR and network coding
is dependent on the number of nodes $N$ in the topology component.
While the gain manifests itself in the throughput of each canonical
topology component, the major benefit is realized in the delay, or
time it takes to complete all flows. }

\textcolor{black}{For purposes of illustration, we now restrict our
analysis to the cases in which we have the restrictive MAC with CSMA,
and we only consider symmetric traffic across each topology component.
Combining equations (\ref{eq:Throughput}) and (\ref{eq:cross_bw1})
through (\ref{eq:x_bw2}), relaxing the integer constraints, and assuming
an equal number of nodes in each set within the {}``X'' topology
component, we take the limit of the throughput for each canonical
topology component:\begin{equation}
\lim_{N\rightarrow\infty}S_{Cross}=\begin{cases}
\frac{m}{m+1} & \textrm{without NC}\\
m & \textrm{with NC}\end{cases}\end{equation}
\begin{equation}
\lim_{N\rightarrow\infty}S_{X}=\begin{cases}
\frac{m}{m+1} & \textrm{without NC}\\
\frac{2m}{m+2} & \textrm{with NC}\end{cases}\end{equation}
It is clear from the above analysis that the gain has a dependency
on the connectivity of the network. As the network becomes more connected,
the interaction between network coding and MPR combine to create gains
that are super-additive.}

\begin{figure}
\begin{centering}
\textcolor{black}{\includegraphics[width=3.5in]{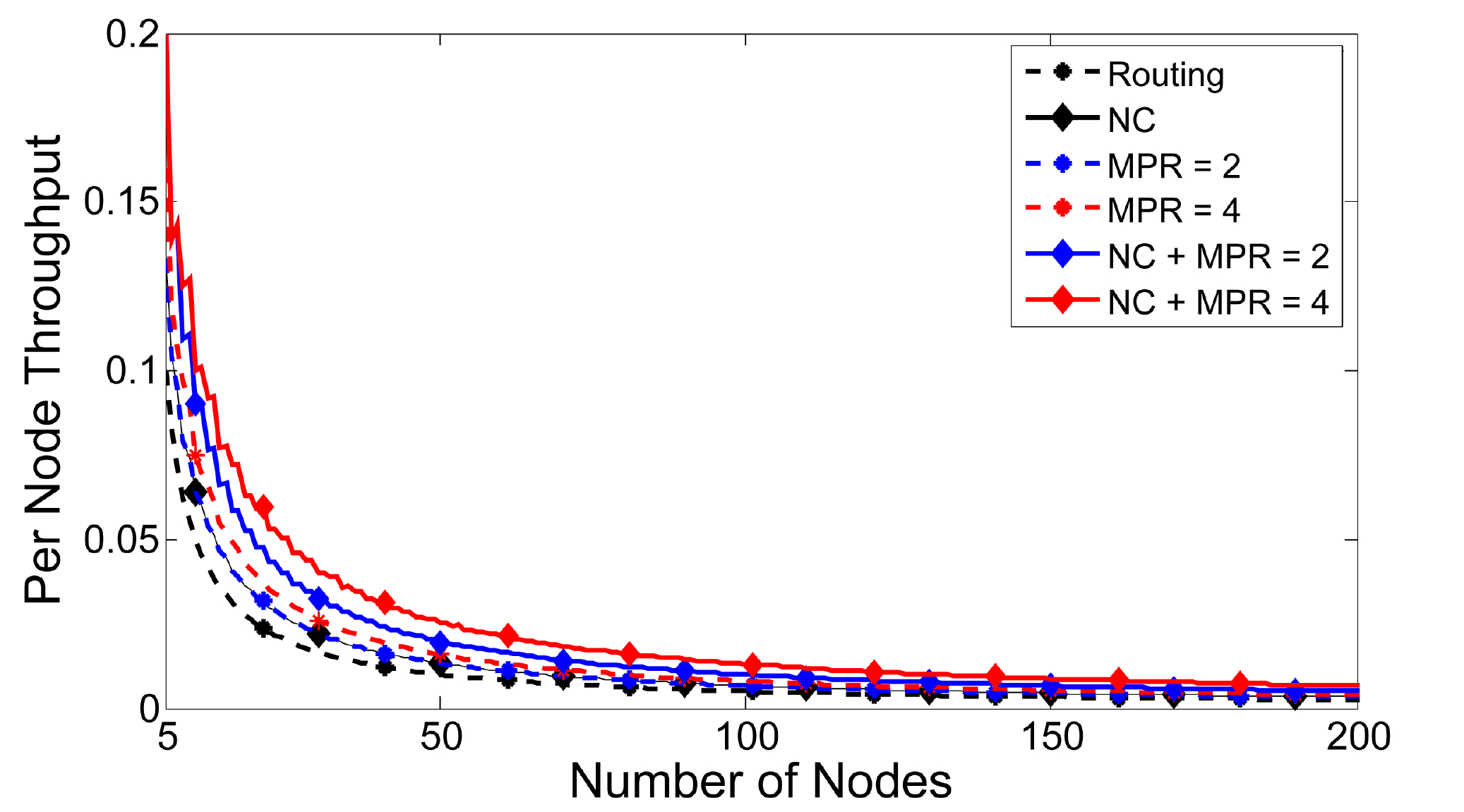}}
\par\end{centering}

\textcolor{black}{\caption{Throughput per node of the {}``X'' topology component for large
N using the improved MAC\label{fig:Throughput-per-node}}
}
\end{figure}

\textcolor{black}{Considering the per node throughput $S_{Node}=s_{j}$
for $j\in[1,N]$, we see that the throughput for both the original
802.11 MAC and improved MAC scales on the order of $\nicefrac{1}{N}$.
Fig. \ref{fig:Throughput-per-node} shows the {}``X'' topology component's
per node throughput, using the improved MAC, as a function of the
number of nodes. As expected, the throughput per node asymptotically
approaches zero as $N$ grows. While there are gains from MPR and
network coding for moderately sized networks, i.e., $N=[5,100]$,
the throughput gains are limited for larger ones.}

\begin{figure}
\begin{centering}
\textcolor{black}{\includegraphics[width=3.5in]{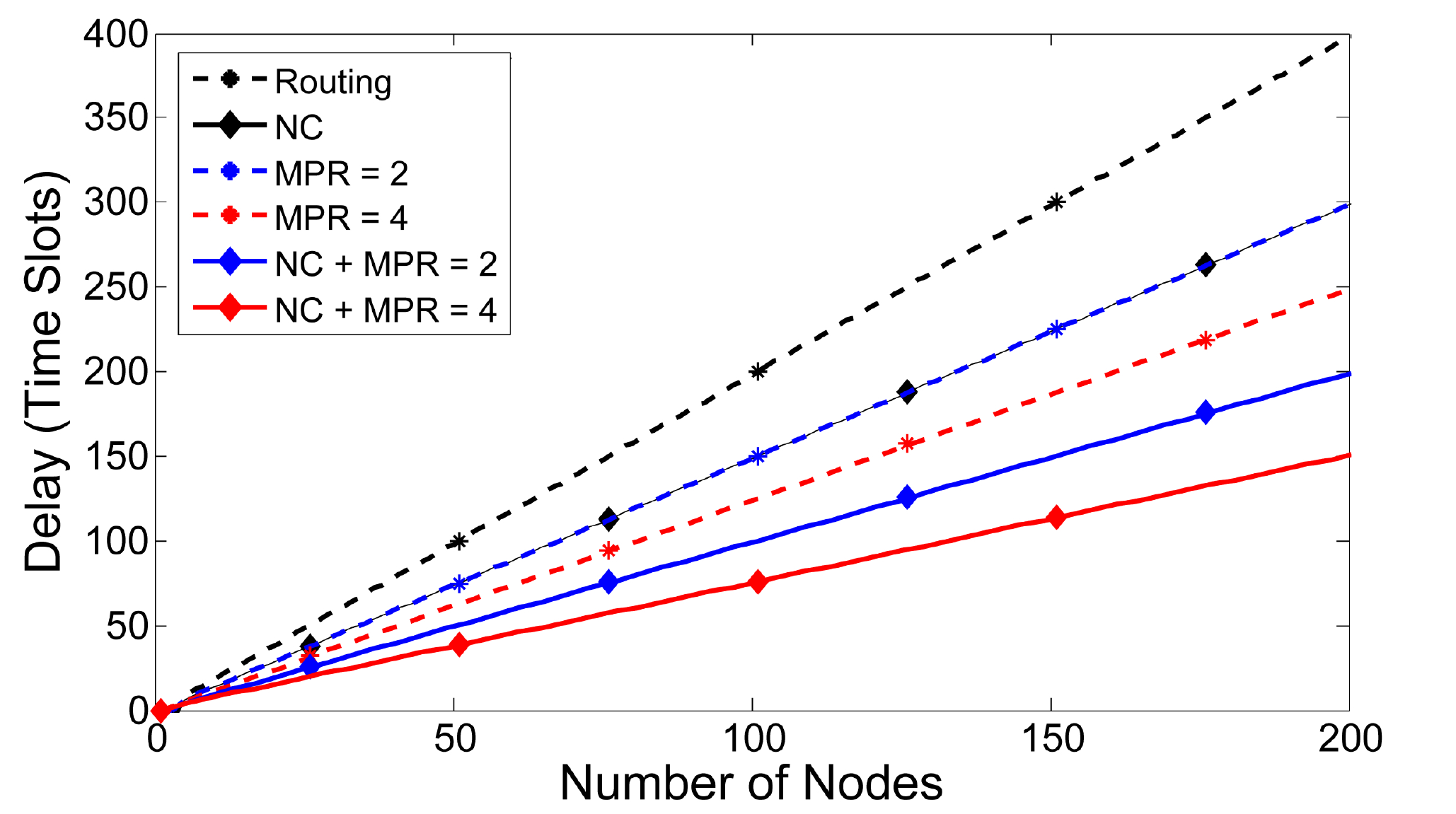}}
\par\end{centering}

\textcolor{black}{\caption{Time to complete all flows if each source has only a single packet
to send using the improved MAC.\label{fig:Delay}}
}

\end{figure}

\textcolor{black}{On the other hand, there are significant gains from
MPR and network coding, while using the improved MAC, when considering
the }\textit{\textcolor{black}{delay}}\textcolor{black}{, or total
time to complete all sessions. When a single packet is at every node,
we determine the time for all packets to reach their intended destinations.
Fig. \ref{fig:Delay} shows the total time to complete all flows within
an {}``X'' topology component, as $N$ grows. It can be easily verified
that the delay gains for the MPR with $m=2$ or $m=4$ and network
coding cases are approximately $2$ and $\nicefrac{8}{3}$ respectively
for large $N$.}%
\textcolor{black}{}%

\section{\textcolor{black}{Conclusion\label{sec:Conclusion}}}

\textcolor{black}{We have provided a lower bound to the gains in total
throughput from MPR and network coding for topology components that
create traffic bottlenecks in large networks. We provided an analysis
of the total throughput and showed that the effectiveness of network
coding is highly dependent on the use of MPR. We have shown that the
combined use of MPR and network coding results in super-additive gains,
rather than just purely additive gains.}

\textcolor{black}{In addition, we evaluated the fairness imposed by
the 802.11 MAC and showed that the coding + MPR gain at saturation
is not maximized. We argued that while the current 802.11 MAC is fair
to }\textit{\textcolor{black}{nodes}}\textcolor{black}{, it is inherently
unfair to }\textit{\textcolor{black}{flows}}\textcolor{black}{{} of
information in multi-hop networks. We further generalized each scenario
for both unicast and broadcast traffic.}

\textcolor{black}{We then used our simple, validated model to design
a new MAC approach using MPR and network coding that allocates channel
resources by providing a greater proportion of resources to bottle-necked
nodes and less to source nodes. The new MAC overcomes the constraints
of the legacy 802.11 MAC and ensures fairness among information }\textit{\textcolor{black}{flows}}\textcolor{black}{{}
rather than }\textit{\textcolor{black}{nodes}}\textcolor{black}{.
Our proposed approach, optimized for networks using network coding
and MPR, shows an increase in the achievable throughput of as much
as 6.3 times the throughput when neither network coding nor MPR is
used in similar networks. Finally, we analyzed the scalability of
the canonical topology components. We showed that the gains provided
by the use of MPR and network coding are highly dependent on the connectivity
of the network and the gains are not necessarily realized in the throughput
but in the time it takes to propagate information for large networks.}

\textcolor{black}{\bibliographystyle{IEEEtran}
\bibliography{COPEMPRBib}
}
\end{document}